\documentstyle[preprint,aps]{revtex}
\begin{document}
\title{Relativistic microscopic description of proton-nucleus scattering
at intermediate energies}
\author{G. Q. Li and R. Machleidt}
\address{Department of Physics, University of Idaho, Moscow ID 83843, USA }
\author{R. Fritz and H. M\"uther}
\address{Institut f\"ur Theoretische Physik, Universit\"at T\"ubingen,
72076 T\"ubingen, Germany}
\author{Y. Z. Zhuo}
\address{Institute of Atomic Energy, P. O. Box 275, Beijing, 102413, China\\
and Institute of Theoretical Physics, P. O. Box 2735, Beijing 100080, China}
\maketitle

\begin{abstract}
 We study elastic proton-nucleus  scattering at intermediate energies.
The nucleon-nucleus optical potential is derived from
the Bonn  nucleon-nucleon  potential and  the
Dirac-Brueckner approach for nuclear matter.
Our calculations, which do not contain any adjustable parameters, yield good
agreement with experiment.
\end{abstract}
\pacs{24.10.Jv,25.40.Cm}
\pagebreak

\section{Introduction}

The microscopic description of the bulk properties of nuclear matter,
finite nuclei and
nuclear reactions in terms of a realistic nucleon-nucleon (NN)
interaction continues to be an interesting topic in nuclear physics.
There are
two aspects to the problem. First, one needs a realistic NN
potential.   This strong interaction should, in principle, be derived from
quantum chromodynamics (QCD).  However, up till now, attempts
to derive the NN interaction from QCD have led only to very crude agreement
with experiment which is unsuitable for
studies of
nuclear many-body systems. Presently, the only quantitative models for the
nuclear force are based on meson exchange.
Well-known examples for such
realistic potentials are the
Bonn \cite{mach1,mach2} and the Paris \cite{paris} potential; we will apply
the former in this work.

The second aspect of the problem is a suitable many-body approach that
is able to  deal with
a realistic NN interaction which has strong short-range components.
The Brueckner approach
and the variational method
have been developed for this purpose. However, when using two-body forces
in these non-relativistic approaches,
one is faced with a fundamental problem:
even the bulk  properties of
nuclear matter cannot be reproduced correctly \cite{mach2,day}.

Encouraged by the success  of the Dirac phenomenology  for
nucleon-nucleus scattering \cite{clar1,clar2}
and the Walecka model for dense
nuclear matter \cite{qhd1,qhd2},
a relativistic extension of the Brueckner method   has been initiated
by Shakin and co-workers \cite{shak1,shak2}, frequently called
the Dirac-Brueckner-Hartree-Fock (DBHF) approach.
This approach has been further  developed by
Brockmann and Machleidt \cite{broc1,broc2} and by ter Haar and Malfliet
\cite{mal1}. Horowitz and Serot have discussed in detail
the basic    aspects involved in the derivation of the relativistic
$G$-matrix \cite{horo1,horo2}.  The common feature of all DBHF results
is that a repulsive relativistic many-body effect is obtained which
is strongly density dependent such that the empirical nuclear matter
properties can be explained, starting from a realistic NN interaction.

There are basically two motivations for the extension of the DBHF
approach to other observables of nuclear physics, e.g.,
nucleon-nucleus scattering. The first motivation is
fundamental: one wants to test if it is possible to describe
nuclear matter,
finite nuclei and nuclear reactions in terms of the same basic  NN
interaction.
Secondly, the relativistic description of intermediate-energy
nucleon-nucleus scattering is by itself
an interesting topic of theoretical nuclear physics.
Its most characteristic feature is the appearance of a strong attractive
scalar and a strong repulsive vector potential, in terms of
which the single-particle motion of the projectile nucleon in the
mean field of the target nucleus can be described using the Dirac equation. In
the Dirac phenomenology developed by Clark and collaborators
\cite{clar2,clar3,clar4}, these potentials
are adjusted to  the nucleon-nucleus scattering data, using
an appropriate number of parameters.
Besides the Dirac phenomenology,  approaches have been
developed in which  nucleon-nucleus scattering is described more
microscopically.
We mention here the relativistic impulse approximation \cite{ria1,ria2,horo3}
and calculations based on the Walecka model \cite{horo4,horo5,zhuo}.

These approaches have some disadvantages.
The Walecka model has no
connection to the free-space  NN interaction.  In the relativistic impulse
approximation, the medium effects are ingored.
Thus it is tempting to develop a method that avoids both of these drawbacks.
The
DBHF calculation based on a
realistic NN interactions  offers this opportunity.
There has been some  work in this direction, e.g., in Refs. \cite{mal1,miya}
the DBHF equations are solved directly in nuclear matter for a nucleon
above the Fermi level.

In this paper we present an alternative approach which combines the
microscopic origin of the DBHF calculation with  the simplicity of the
Walecka model, such that an investigation of
intermediate-energy nucleon-nucleus scattering is feasible.

The direct extension of the DBHF calculation from nuclear
matter to finite nuclei and nuclear reactions is not trivial. More
practically, one can parameterizes the DBHF results
for nuclear matter in terms of an effective Lagrangian  which leads
to the same predictions for nuclear matter
as the original DBHF
calculation. The effective Lagrangian, with its parameters derived
from the realistic NN interaction, can then be used to investigate the
properties of finite nuclei and nuclear reactions. Different schemes for this
parameterization have been proposed \cite{muth,marc,pol,broc3}. We use
in the present work a scheme similar to the one which was
recently suggested  by Brockmann and
Toki \cite{broc3}: The DBHF results for nuclear matter obtained from the Bonn
potential are parameterized
in terms of an effective $\sigma$-$\omega$ Lagrangian allowing for
density-dependent meson-nucleon coupling constants.

We discuss in Section II the details of our calculation. The results
and a discussion are presented in Section III, Finally a short summary
is given in Section IV.

\section{Details of the Calculation}

The one-boson-exchange (OBE) Bonn potential \cite{mach2,mach3}
used in this work is constructed in terms of the
Thompson \cite{thom} equation.
The kernel of this integral equation is  the
sum of one-meson-exchange amplitudes of six nonstrange  bosons
with given mass and coupling.
Pseudovector (derivative) coupling is used for
pseudoscalar mesons ($\pi $ and $\eta $).
A  form factor of monopole type
is applied to each meson-nucleon vertex which simulates the short-range
dynamics of quark-gluon nature.
See Ref. \cite{mach2,mach3} for details about the OBE Bonn potential and its
description of the two-nucleon system.

As in conventional Brueckner theory, the basic quantity in the
DBHF  approach is a reaction matrix, $\tilde G$, which satisfies
the in-medium Thompson equation.
The nuclear matter properties are then obtained  from this effective in-medium
two-body interaction. Using  the Bonn A potential
\cite{mach2}, the
DBHF calculation predicts that nuclear matter saturates at
a density
$\rho $=0.185 fm$^{-3}$ and an energy  per nucleon of --15.6 MeV,
which is in good agreement with empirical information.
More   results and discussion
concerning the  properties of nuclear matter as obtained in the DBHF
approach with the Bonn potential have been presented in Refs.
\cite{mach2,broc2,li1}.
In this work, we use the Bonn A potential \cite{bonna}.

As proposed by Brockmann and Toki \cite{broc3}, the
DBHF results for nuclear matter can be parameterized by an
effective Lagrangian, in analogy to the $\sigma$-$\omega$ model of
Walecka:
$${\cal L}=\overline \psi [i\gamma _\mu\partial ^\mu -m-g_\sigma (\rho )
\phi _\sigma -g_\omega (\rho )\gamma _\mu \phi _\omega ^\mu ]\psi
+{1\over 2}(\partial ^\mu \phi _\sigma )^2$$
$${-{1\over 2}m^2_\sigma\phi _\sigma ^2
-{1\over 4}(\partial _\mu \phi _\omega ^\nu -\partial _\nu\phi _\omega ^\mu )^2
+{1\over 2}m^2_\omega \phi _\omega ^{\mu 2}}\eqno (1)$$
where $\psi$ is the nucleon field, while $\phi _\sigma$ and $\phi _\omega ^\mu$
are  effective sigma and
omega fields, respectively. The masses of the  sigma and omega
mesons, $m_\sigma$ and $m_\omega$, resp.,
are kept fixed at their values in  free-space scattering
($m_\sigma = 550$ MeV, $m_\omega = 782.6$ MeV).
However,  the coupling constants of these effective mesons,
$g_\sigma$ and $g_\omega$,
depend on the baryon density $\rho$. They are
determined from the DBHF results for
nuclear matter.

The (complex) nucleon self-energy produced by the effective meson exchanges in
nuclear matter can, in general, be written as
$${\Sigma (k_\mu )=\Sigma _S(k_\mu )+\gamma ^0\Sigma _0(k_\mu )+
{\bf \gamma \cdot k}\Sigma _V(k_\mu )=V+iW}\eqno (2)$$
where $\Sigma _S$, $\Sigma _0$ and $\Sigma _V$ denote the scalar
component, the time-like part of the vector component and the space-like
part of the vector component of the nucleon self-energy, respectively.

In our notation, $k_\mu$ stands for all four components of the momentum
four-vector, and $k\equiv |{\bf k}|$.

Treating the effective coupling constants
locally as numbers and calculating
in the relativistic Hartree-Fock
approximation (see Fig. 1a and 1b for
the corresponding Feynman diagrams), we obtain the
explicit expressions for the real part of the nucleon self-energy
\cite{qhd2,horo4,zhuo}
$${V_S(k,\rho)=-({g_\sigma (\rho )\over m_\sigma })^2\rho _S+
{1\over 16\pi ^2k}\int ^{k_F}_0dqq{m^*_q(\rho)\over E_q^*(\rho)}
\{g_\sigma ^2(\rho )\Theta _\sigma
(k ,q)-4g_\omega ^2(\rho )\Theta _\omega (k ,q)\}}\eqno (3)$$
$${V_0(k,\rho)=({g_\omega (\rho )\over m_\omega} )^2\rho +
{1\over 16\pi ^2k}\int ^{k_F}_0dqq\{g_\sigma ^2(\rho )\Theta _\sigma
(k,q)+2g_\omega ^2(\rho )\Theta _\omega (k,q)\}}\eqno (4)$$
$${V_V(k,\rho)=
-{1\over 8\pi ^2k^2}\int ^{k_F}_0dqq {q^*\over E^*_q(\rho)}
\{g_\sigma ^2(\rho )\Phi
_\sigma
(k,q)+2g_\omega ^2(\rho )\Phi _\omega (k,q)\}}\eqno (5)$$
where
$${\rho _S=4\int _0^{k_F}{d^3q\over (2\pi )^3}{m^*_q(\rho)\over
E^*_q(\rho)}}
$$
$${\rho =4\int _0^{k_F}{d^3q\over (2\pi )^3}}$$
and
$$\Theta _i (k,q)=ln|{(k+q)^2+m_i^2-(q_0-k_0)^2\over
(k-q)^2+m_i^2-(q_0-k_0)^2}|
$$
$$\Phi (k,q)={k^2+q^2+m_i^2-(k_0-q_0)^2\over 4kq}\Theta _i(k,q)-1,~~i=\sigma ,
\omega$$
The effective mass $m^*$, the kinetic momentum ${\bf k}^*$ and
the single particle energy $k_0$ are given by
$$m^*_k(\rho)=m+V_S(k,\rho)$$
$${\bf k}^*={\bf k}(1+V_V(k))$$
$$k_0=E^*_k+V_0=({\bf k}^{*2}+m^{*2}_k)^{1/2}+V_0$$
and similarly for $q$.

The first term of $V_S$ and $V_0$ is  the
momentum-independent Hartree contribution (see Fig. 1a), while the other terms
are the Fock contributions (see Fig.~1b) which depend on $k$.
A very small space-like component of the vector potential
is due entirely to
Fock diagrams.

In order to determine the effective coupling constants from these expressions,
we drop the space-like component of the vector potential (Eq. (5)) and
calculate $V_S$ (Eq. (3))
and $V_0$ (Eq. (4)) for $k=k_{F}$, which are then  identified  with
the scalar and vector potentials obtained in the DBHF calculation
(see Table VII of Ref.~\cite{broc2}). This is a
reasonable assumption since the space-like component of the vector
potential is very small and the potentials are only very weakly momentum
dependent \cite{Mac86}. Note that these approximations are used only for the
determination of the effective coupling constants; in the calculation of
the optical potential, the space-like part of the vector potential (Eq. (5))
is taken into account (see Eq. (10) below).
The momentum $k$ in Eqs. (3)-(5) is related to the
incident energy of the projectile (see Eq. (11) below) through $k_0=E$.
Both effective coupling constants drop
with increasing density \cite{li2,fritz}.

The lowest-order contribution to the imaginary part of the nucleon
self-energy is obtained from the fourth-order (counting the
meson-nucleon vertices) Feynman diagrams which are
characterized by two-particle-one-hole
($2p1h$) intermediate states (see Fig. 1c). The nucleon lines in these
Feynman diagrams are described by dressed nucleon propagators, which
corresponds to performing the calculation on the Hartree-Fock
ground state. The imaginary part of nucleon self-energy due to the fourth-order
Feynman diagrams can be expressed as \cite{horo5,zhuo}

$${W_S(k,\rho)=
{1\over 8\pi ^2k}\int ^{E'}_0dq_0\int ^{k+p}_{|k-p|}dq q[g_\sigma ^2(\rho )
m^*\Delta ^2_\sigma (q){\rm Im}\pi _\sigma (q)}$$
$${+g_\omega ^2(\rho )m^*\Delta _\omega ^2(q)g_{\mu\lambda}{\rm Im}\pi _\omega
^{\lambda\mu}(q)+2g_\sigma (\rho )g_\omega (\rho )(E^*_{\bf k}-{1\over 2}q_0)
{q^2_\mu\over q^2}\Delta _\sigma (q)\Delta _\omega (q){\rm Im} \pi ^0_{\sigma
\omega}(q)]}\eqno (6)$$
$${W_0(k,\rho)={
1\over 8\pi ^2k}\int ^{E'}_0dq_0\int ^{k+p}_{|k-p|}dq q[g_\sigma ^2(\rho )
\Delta ^2_\sigma (q)(E_{\bf k}^*-q_0){\rm Im}\pi _\sigma (q)}$$
$${-g_\omega ^2(\rho )\Delta _\omega ^2(q)
[(E^*_{\bf k}-q_0)g_{\mu\lambda}{\rm Im}\pi _\omega
^{\lambda\mu}(q)+(2E^*_{\bf k}-q_0){q^2_\mu\over q^2}{\rm Im}\pi ^{00}_\omega
(q)]}$$
$${-2g_\sigma (\rho )g_\omega (\rho )m^*
\Delta _\sigma (q)\Delta _\omega (q){\rm Im} \pi ^0_{\sigma
\omega}(q)]}\eqno (7)$$
$${W_V(k,\rho)=-{1\over 8\pi ^2k^3}\int ^{E'}_0dq_0\int ^{k+p}_{|k-p|}dq
q[g_\sigma ^2(\rho )
\Delta ^2_\sigma (q)(k^2+{1\over 2}q^2_\mu -E^*_{\bf k}
q_0)){\rm Im}\pi _\sigma (q)}$$
$${+g_\omega ^2(\rho )\Delta _\omega ^2(q)
[-(k^2+{1\over 2}q^2_\mu -E^*_{\bf k}q_0)g_{\mu\lambda}{\rm Im}\pi _\omega
^{\lambda\mu}(q)+{q_0q^2_\mu\over q^4}(2E_{\bf k}^*q_0-q_\mu ^2)({1\over 2}
q_0-E^*_{\bf k}){\rm Im}\pi ^{00}_\omega
(q)]}$$
$${-g_\sigma (\rho )g_\omega (\rho )m^*{q_0\over q}(2k_0^*q_0-q_\mu ^2)
\Delta _\sigma (q)\Delta _\omega (q){\rm Im} \pi ^0_{\sigma
\omega}(q)]}\eqno (8)$$
where $q$ is the magnitude of the three-vector $q$, while
$$ m^* = m^*_k(\rho)$$
$$E'=E^*_{\bf k}-(k_F^2+m^{*2})^{1/2}$$
$$p=((E^*_{\bf k}-q_0)^2-m^{*2})^{1/2}$$
$$q_\mu ^2=q_\mu q^\mu$$
$$\Delta _i(q)={1\over q_\mu ^2-m_i^2}, ~~i=\sigma , ~\omega$$
$\pi(q)$ are meson
polarization insertions; their imaginary parts are given below
$${\rm Im}\pi _\sigma (q)=-{g^2_\sigma (\rho )\over \pi q}(m^{*2}-{1\over 4}
q^2_\mu )(E^*_F-E^*)$$
$${\rm Im}g_{\mu\lambda}\pi _\omega ^{\lambda\mu}(q)=
-{g^2_\omega (\rho )\over \pi q}
(m^{*2}+{1\over 2}
q^2_\mu )(E^*_F-E^*)$$
$${\rm Im}\pi _\omega ^{00}(q)=-{g^2_\omega (\rho )\over \pi q}
({1\over 3}(E^{*3}_F-E^{*3})+{1\over 2}(E^{*2}_F-E^{*2})q_0
+{1\over 4}q_\mu ^2(E_F^*-E^*))$$
$${\rm Im }\pi ^0_{\sigma\omega} (q)= -{g_\sigma (\rho )g_\omega (\rho )\over
2\pi q}m^*((E^{*2}_F-E^{*2})+(E^*_F-E^*)q_0)$$
where
$$E^*=max\{E_r,E_F^*-q_0,m^*\}$$
$$E_r=-{1\over 2}q_0+{1\over 2}q(1-4m^*/q^2_\mu )^{1/2},~~E^*_F=(k_F^2+m^{*2})
^{1/2}$$

The derivation of the nucleon self-energy
(Eqs. (3)-(8)) from the Walecka model has been discussed
in detail in Refs. \cite{horo4,horo5}. For the effective Lagrangian (Eq. (1))
used in the present work, the expressions for the nucleon self-energy are
the same, only with the coupling constants of the sigma- and
omega-exchanges replaced by those determined from the DBHF
approach which are now density dependent.

The same fourth-order diagrams, which yield the imaginary part also
give rise to a contribution to the real part of the self-energy, which
should be added to the Hartree-Fock term. Note, however, that the
Hartree-Fock contribution has been derived from a G-matrix description
of nuclear matter and therefore already contains these terms in a
certain approximation. In order to avoid any double counting no
contributions of the fourth order diagram to the real part have been
considered.

The nucleon self-energy in a finite nucleus
is obtained by means of the local density approximation,
in which the spatial dependence of the microscopic optical potential is
directly related to the density of the target nucleus.
For a self-consistent description of nucleon-nucleus scattering
(by self-consistency we mean that there are no free parameters after
the NN potential has been selected), the
target density should  also be
determined from the effective Lagrangian, Eq. (1).
In this paper, we use  the density
determined in the relativistic  density-dependent Hartree-Fock (RDHF)
calculation with
the effective Lagrangian of Eq. (1) \cite{fritz}.
For comparison, we also use the density determined in a relativistic
density-dependent Hartree (RDH)
calculation \cite{fritz}, as well as the density determined in the
relativistic Hartree (RH) calculation with the Walecka model \cite{horo33}.
As shown in Ref. \cite{fritz}, the RDH results for the
binding energy and root-mean-square radius of $^{40}$Ca are in good agreement
with experiment, while the RDHF results slightly underestimates them.

The Dirac equation for the single-particle motion of the projectile
nucleon in the mean field of the target nucleus
can be written as
$${[{\alpha\cdot}{\bf k}+\beta (m+U_S)+U_V+V_C]\psi =E\psi}\eqno (9)$$
with
$${U_S={\Sigma _S-m\Sigma _V\over 1+\Sigma _V},~~U_V={\Sigma _0+E\Sigma _V\over
1+\Sigma _V}}\eqno (10)$$
where $E$ is the energy of the projectile in the center-of-mass
(c.m.) system
of projectile and nucleus, which is related to the incident energy
$T_{lab}$ through
$${E=E_{p,c.m.}={m^2+m_T(m+T_{lab})\over ((m+m_T)^2+2m_TT_{lab})^{1/2}}}\eqno
(11) $$
with $m$ and $m_T$ the mass of the projectile and target, respectively.
$V_C$ is the Coulomb field which we treat in the same way as in Refs.
\cite{horo3,horo33}.

In order
to calculate the experimental observables, the Dirac equation is usually
converted into a Schr\"odinger-equivalent equation.
The Dirac equation for the four-component spinor $\psi $ is equivalent to two
coupled equations for the large (upper) and small (lower) two-component
spinors.
One can eliminate the small component of the Dirac spinor in a
standard way and obtain a Schr\"odinger-equivalent equation for the large
component of the Dirac spinor \cite{clar2,horo3,horo33}
$${\left[{{\bf k}^2\over 2E}+U_{eff}(r)+V_C(r)+U_{s.o.}(r){\bf \sigma \cdot l}
\right]
\phi ({\bf r})={E^2-m^2\over 2E}\phi ({\bf r})}\eqno (12)$$
where $U_{eff}$ and $U_{s.o.}$ are the central  and the spin-orbit part
of the Schr\"odinger-equivalent potential, which is known as the
nucleon-nucleus optical potential in non-relativistic approaches.
The explicit expressions for $U_{eff}$
and $U_{s.o.}$ are
$${U_{eff}=U_V+{1\over 2E}[U_S(2m+U_S)-(U_V+V_C)^2+U_{Darwin}]}\eqno (13)$$
$${U_{s.o.}=-{1\over 2ErD(r)}{dD(r)\over dr}}\eqno (14)$$
with
$${U_{Darwin}={3\over 4}\left[{1\over D(r)}{dD(r)\over dr}\right]^2-{1\over
rD(r)}{dD(r)\over dr}-{1\over 2D(r)}{d^2D(r)\over dr^2}}$$
$$D(r)=m+E+U_S-U_V-V_C$$

\section{Results and Discussion}

We show in Fig. 2 the density of $^{40}$Ca determined in a RDHF (solid
curve)
and RDH (dashed curve) calculation~\cite{fritz}
 using  the effective Lagrangian
of Eq. (1); they are compared with the density (dotted curve) as obtained
from a
relativistic Hartree (RH) calculation applying  the Walecka model (QHD-I)
\cite{horo33}. Note that the RDHF calculation slightly underestimates the
root-mean-square radius of $^{40}$Ca \cite{fritz}.

We show in Fig. 3 the scalar potential $U_S$ and vector potential $U_V$
for p+$^{40}$Ca scattering
obtained in the present calculation. The RDHF density is used
in the local density approximation. The real parts of these potentials
are only weakly energy dependent; in the interior of the nuclei, they are
about $-420$ and $+320$ MeV, respectively. The overwhelming part of these real
potentials comes from the energy independent Hartree contribution; the
energy dependent Fock contribution plays only a minor role. On the
other hand the  imaginary part of the scalar and vector potential
depends of course very strongly on the energy;
both increase in magnitudes with increasing
energy. Note that the radial shapes of the real Dirac potentials follow
essentially that of the target density (cf.  solid curve in Fig. 2).

With the nucleon self-energy determined from the effective Lagrangian
which is related to the realistic NN interaction
used in the DBHF calculation,
we can obtain the Schr\"odinger-equivalent potential through
Eqs. (13) and (14) which is then used in Eq. (12) to calculate the
experimental observables~\cite{horo33}.
The Schr\"odinger-equivalent potentials,
corresponding to the Dirac potentials in Fig. 3,  are shown in Fig. 4.
As can be seen from the figure,
the real part of the central potential
is strongly energy dependent. At low incident energy
($T_{lab}$=150) the real part of the central potential is
attractive throughout the whole radial space and has a longer range. With
the increase of the incident energy, the real part of the central
potential becomes less attractive and is  repulsive in the interior of
the nucleus when $T_{lab}$=300 MeV. A typical pocket of attraction
is observed at the nuclear surface at this energy.
At even higher energy ($T_{lab}$=450 MeV), the real central potential
is repulsive in the whole radial space.
The imaginary part of the central potential  also shows a strong energy
dependence.
The imaginary central potential is negative at all energies;
its magnitude increases while its range
decreases with increasing  energy.

The important feature of the relativistic approach is that the spin-orbit
potential arises naturally from the coherent sum of the
contributions from  the scalar and  vector potential.
While the real part of the spin-orbit potential is attractive,
its
imaginary part is mostly positive.
The real part of the spin-orbit potential
decreases (in magnitude) with  increasing incident energy, whereas its
imaginary part increases with the energy.
The fluctuations of the spin-orbit potential in the interior of the nucleus
are due to fluctuations in the target  density (cf.  solid curve
in Fig. 2).

The results for differential cross sections and
analysing powers in elastic  p+$^{40}$Ca scattering at $T_{lab}$=300 and 400
MeV
are shown in Fig. 5 and Fig. 6, respectively. In these figures,
the solid and dashed curves are the results of our calculations
with RDHF and RDH densities, respectively, while for the dotted curves
the RH density
from the Walecka model (QHD-I) \cite{horo33} is used.
The experimental data
are from Ref. \cite{olsen}.

For full self-consistency of our calculations, we must
use the RDHF density since it is based upon the same
effective coupling constants as our optical
potential.
With the RDHF density, our results
(solid curves) are in reasonable agreement with experiment.
The agreement is particularly good at small angles ($\theta _{c.m.}\le
15^o$). At large angles, we overestimate the experimental differential
cross section. The oscillations in the differential cross section and
analysing power seem to have a larger period than the experimental
data show (note that there is a correspondence between the minima in the
differential cross section and the dips in the analysing power).
Since our calculations are parameter free, the quality of agreement
with the experimental data
is remarkable. The remaining discrepancies are probably mainly due to the
fact that
the target density as obtained in the RDHF calculation underestimates the
root-mean-square radius \cite{fritz} (see discussion below).

When using the RDH density, which is in better agreement with the
experimental root-mean-square radius \cite{fritz}, our results
(dashed curves in Figs.~5 and 6) improve; and
there is further improvement
when the RH density (dotted curves) is used.
Thus, the accurate reproduction of the nucleon density
of finite nuclei is probably the most important
outstanding problem for the self-consistent
description of nuclear structure and nuclear reactions.

\section{Summary}
In this paper, we have continued our efforts of obtaining a
parameter-free and selfconsistent description of
nuclear matter, finite nuclei, and nuclear reactions in terms
of one  realistic NN interaction.
Here, we have studied p+$^{40}$Ca scattering in
Dirac dynamics.
In order to carry out a systematic
study, we parameterized the DBHF results for nuclear matter
in terms of a simple effective Lagrangian consisting of an attractive
sigma-exchange and a repulsive  omega-exchange. The
coupling constants of these effective mesons are density dependent and
are determined from the DBHF results for nuclear matter.

The nucleon self-energy in nuclear matter is then calculated based on
the effective Lagrangian up to the fourth order Feynman diagrams. The
nucleon-nucleus optical potential in finite nuclei
is related to the self-energy
in nuclear matter by means of the local density approximation,
where the target density is determined by the same effective Lagrangian
as the nucleon self-energy. Furthermore,
the Dirac equation describing the single-particle motion of
the projectile in the mean field of the target is converted to the
Schr\"odinger equivalent equation which is solved to obtain the
observables.

With the optical potential thus determined,
we have calculated observables of p+$^{40}$Ca  scattering at intermediate
energies.
The predictions by our model for the elastic differential  cross section and
the analyzing power
were compared with
the experimental data.
The agreement between
our parameter-free calculations  and  experiment is remarkable.
At large scattering angles,
our results  for the differential cross section overestimate
the experimental data.
This can be attributed, in part, to small deficiencies in the
densities as obtained in the RDHF calculation.

\vskip 1cm
{\bf Acknowledgement}.  One of the authors (GQL) likes to thank
Prof. W. C. Olsen for the experimental data file.
This work is supported in part by the U.S. National
Science Foundation under Grant No. PHY-9211607 and by the Idaho
State Board of Education.
\pagebreak

\pagebreak

\centerline{\bf Figure Captions}
\vskip 0.6cm

{\bf Fig. 1}: Feynman diagrams for the calculation of the nucleon self-energy
in nuclear matter. (a) Hartree contribution , (b) Fock contribution ,
(c) fourth-order
contribution.

{\bf Fig. 2}: Nucleon density of $^{40}$Ca as obtained in  RDHF (solid curve)
and
RDH (dashed curve) calculations with the effective Lagrangian, Eq. (1)
\cite{fritz}; the dotted
curve  represents a RH calculation with the Walecka (QHD-I)
model~\cite{horo33}.

{\bf Fig. 3}: Scalar and vector potentials in p+$^{40}$Ca
scattering at 150 (solid curves), 300 (dashed curves) and 450
MeV (dotted curves). The RDHF density is used in
local density approximation.

{\bf Fig. 4}: The Schr\"odinger-equivalent potential in p+$^{40}$Ca
scattering at 150 (solid curves), 300 (dashed curves) and
450 MeV (dotted curves). The RDHF density is used in  local
density approximation.

{\bf Fig. 5}: Elastic differential cross section and  analyzing power
in p+$^{40}$Ca scattering at $T_{lab}$=300 MeV.
The solid and dashed  curves are the predictions by the present calculations
with RDHF (solid curve in Fig. 2) and RDH
(dashed curve in Fig. 2)
densities, respectively, while the dotted curves
are the results using  the RH density from
the Walecka (QHD-I) model     (dotted curve in Fig. 2).
The experimental data
are from Ref.  \cite{olsen}

{\bf Fig. 6}: Same as Fig. 3, but at $T_{lab}$=400 MeV.


\begin{thebibliography}{99}
\bibitem{mach1} R. Machleidt, K. Holinde and Ch. Elster, Phys. Rep.
{\bf 149}, 1 (1987).
\bibitem{mach2} R. Machleidt, Adv. Nucl. Phys. {\bf 19}, 189 (1989).
\bibitem{paris} M. Lacombe, B. Loiseau, J. M. Richard, R. Vinh Mau, J. Cote,
P. Pires and R. de Tourreil, Phys. Rev. C. {\bf 21}, 861 (1980).
\bibitem{day} B. D. Day and R. B. Wiringa, Phys. Rev. C. {\bf 32}, 1057 (1985).
\bibitem{clar1} L. G. Arnold, B. C. Clark and R. L. Mercer,  Phys.
Rev. C. {\bf 19}, 917  (1979).
\bibitem{clar2} B. C. Clark, in {\it Relativistic Dynamics and Quark-Nuclear
Structure}, edited by M. B. Johnson and A. Picklesimer (Wiley, New York, 1986).
\bibitem{qhd1} J. D. Walecka, Ann. Phys. (N.Y.) {\bf 83}, 491 (1974).
\bibitem{qhd2} B. D. Serot and J. D. Walecka, Adv. Nucl. Phys.
{\bf 16}, 1 (1986).
\bibitem{shak1}M. R. Ansatasio, L. S. Celenza, W. S. Pong and
C. M. Shakin, Phys. Rep. {\bf 100}, 327 (1983).
\bibitem{shak2} L. S. Celenza and C. M. Shakin, Relativistic Nuclear Physics:
Theories of Structure and Scattering, Lecture Notes in Physics, Vol. 2
(World Scientific, Singapore, 1986).
\bibitem{broc1} R. Brockmann and R. Machleidt, Phys. Lett. {\bf 149B}, 283
(1984).
\bibitem{broc2} R. Brockmann and R. Machleidt, Phys. Rev. C
{\bf 42}, 1965 (1990).
\bibitem{mal1} B. ter Haar and R. Malfliet, Phys. Rep. {\bf 149}, 207 (1987).
\bibitem{horo1} C. J. Horowitz and B. D. Serot, Phys. Lett. {\bf 137B},
287 (1984).
\bibitem{horo2} C. J. Horowitz and B. D. Serot, Nucl. Phys. {\bf A464},
613 (1987).
\bibitem{clar3} S. Hama, B. C. Clark, E. D. Cooper, H. S. Sherif and R. L.
Mercer, Phys. Rev. C {\bf 41}, 2737 (1990).
\bibitem{clar4} E. D. Cooper, S. Hama, B. C. Clark and R. L. Mercer,
Phys. Rev. C. {\bf 47}, 297 (1993).
\bibitem{ria1} J. A. McNeil, J. R. Shepard and S. J. Wallace, Phys.
Rev. Lett. {\bf 50}, 1493  (1983)
\bibitem{ria2} S. J. Wallace, Ann. Rev. Nucl. Part. Sci. {\bf 37}, 267 (1987).
\bibitem{horo3} D. P. Murdock and C. J. Horowitz, Phys. Rev. C. {\bf 35}, 1442
(1987).
\bibitem{horo4} C. J. Horowitz, B. D. Serot, Nucl Phys.
{\bf A399}, 529 (1983).
\bibitem{horo5} C. J. Horowitz, Nucl. Phys. {\bf A412}, 228 (1984).
\bibitem{zhuo} Z. Y. Ma, P. Zhu, Y. Q. Gu and Y. Z. Zhuo,
Nucl. Phys. {\bf A490}, 619 (1988).
\bibitem{miya} Y. Miyama, Phys. Lett. {\bf 215B}, 602 (1988).
\bibitem{muth} H. Elsenhans, H. M\"uther and R. Machleidt, Nucl. Phys.
{\bf A515}, 715 (1990).
\bibitem{marc} S. Marcos, M. Lopez-Quelle and N. Van Giai, Phys. Lett. {\bf
257B}, 5 (1991).
\bibitem{pol} S. Gmuca, Nucl. Phys. {\bf A547}, 447 (1992).
\bibitem{broc3} R. Brockmann and H. Toki, Phys. Rev. Lett. {\bf 68}, 340
(1992).
\bibitem{mach3} R. Machleidt, in {\it Computational Nuclear Physics},
Vol. II,  S. E. Koonin, K. Langanke and A. Maruhn, eds. (Springer, New
York, 1993) 1.
\bibitem{thom} R. H. Thompson, Phys. Rev. D {\bf 1}, 710 (1970).
\bibitem{li1} G. Q. Li, R. Machleidt and R. Brockmann, Phys. Rev.
{\bf C45}, 2782 (1992).
\bibitem{bonna} The Bonn A potential, which is applied in the present
work, is defined in Table A.2 of Ref. \cite{mach2}.
\bibitem{Mac86} R. Machleidt, in {\it Relativistic Dynamics and
Quark-Nuclear Physics}, M. B. Johnson and A. Picklesimer, eds.
(Wiley, New York, 1986) p. 71.
\bibitem{li2} G. Q. Li, R. Machleidt and Y. Z. Zhuo, Phys. Rev. C,
submitted.
\bibitem{fritz} R. Fritz, H. M\"uther and R. Machleidt, Phys. Rev.
Lett. {\bf 71}, 46 (1993).
\bibitem{horo33} C. J. Horowitz, D. P. Murdock, and B. D. Serot,
in {\it Computational Nuclear Physics}, Vol.~I, K. Langanke,
A. Maruhn, and S. E. Koonin, eds. (Springer, New York, 1991).
\bibitem{olsen} D. A. Hutcheon {\it et al.,}  Nucl. Phys. {\bf A483},
429 (1988); W. C. Olsen, private communication.
\end{thebibliography}
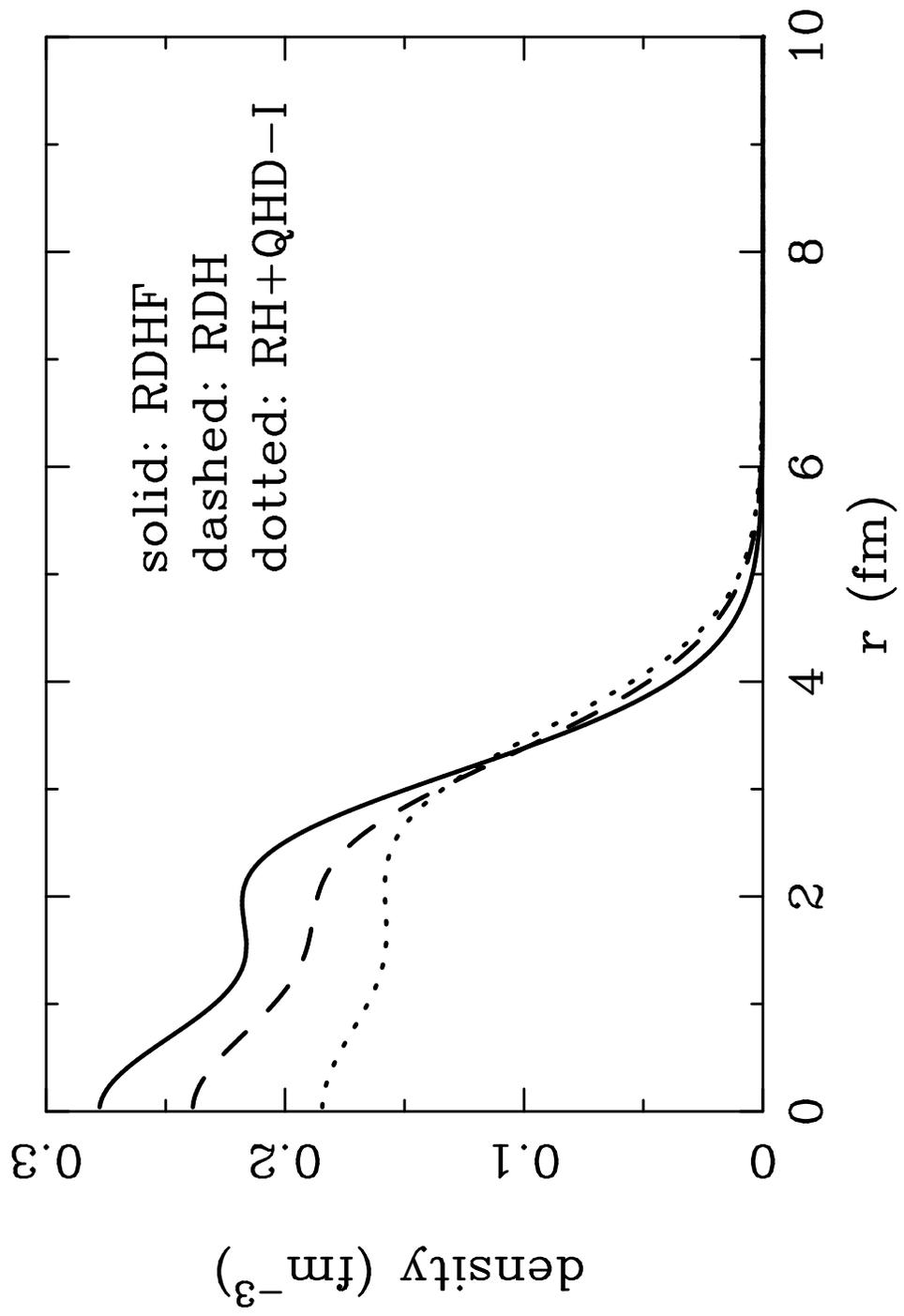
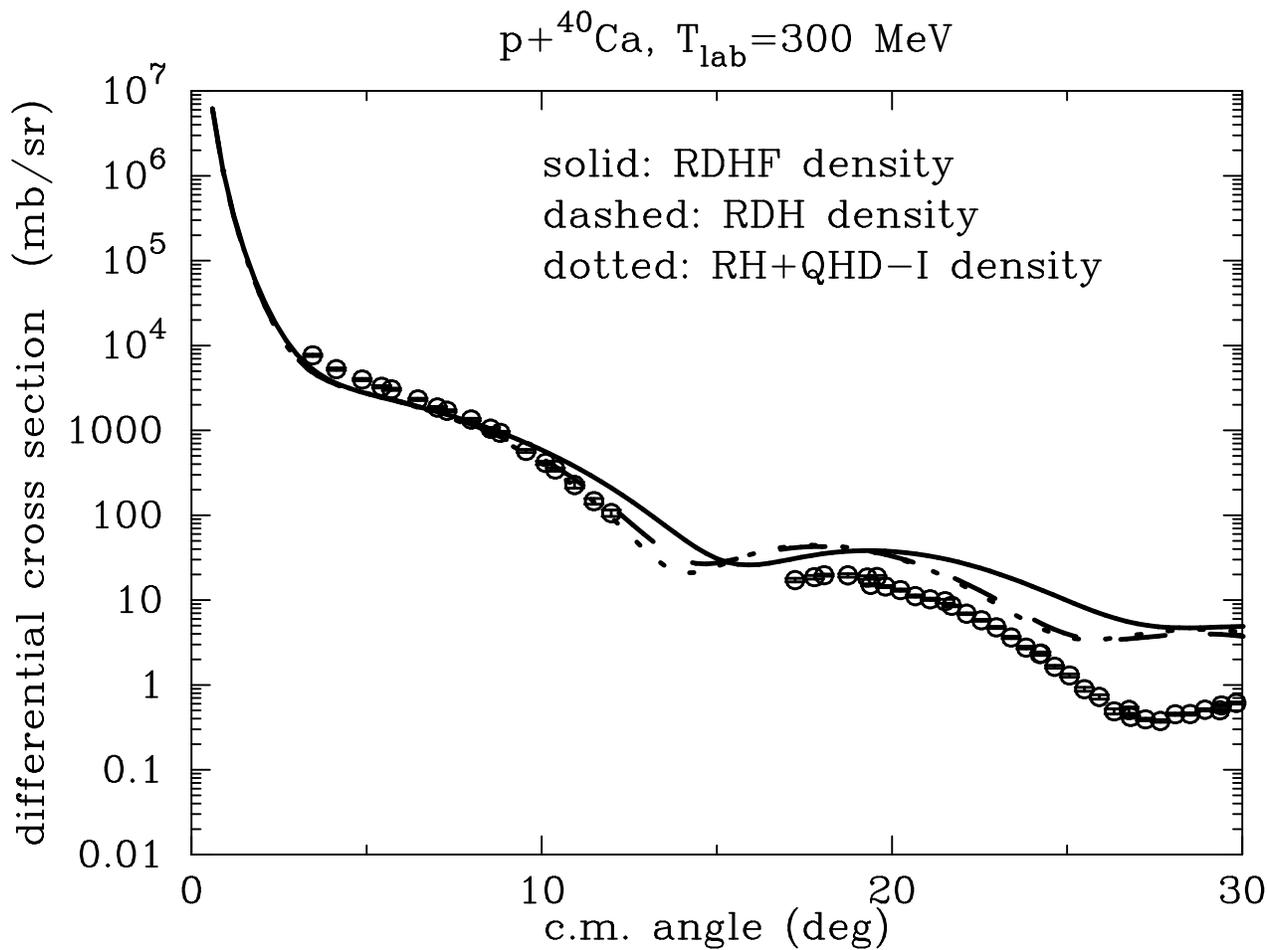
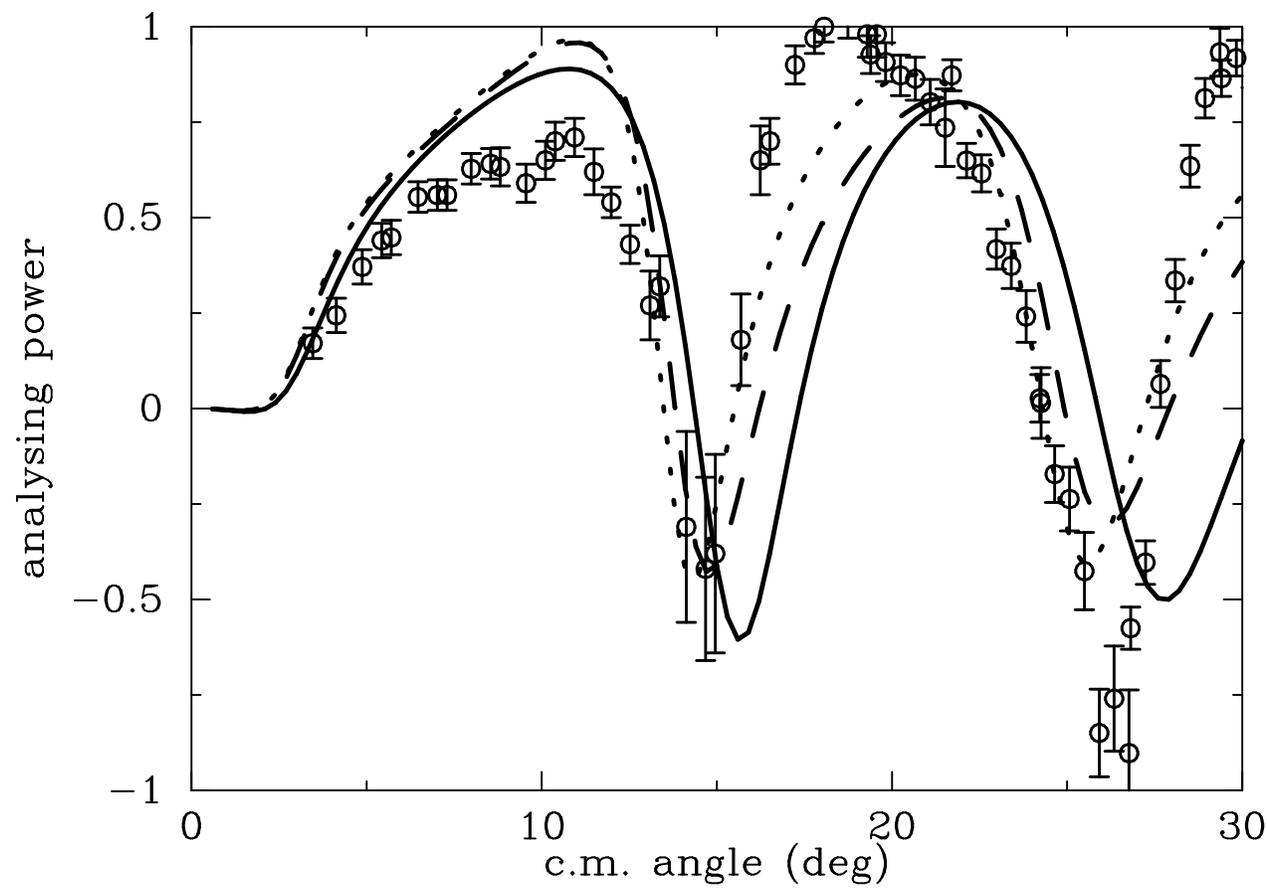
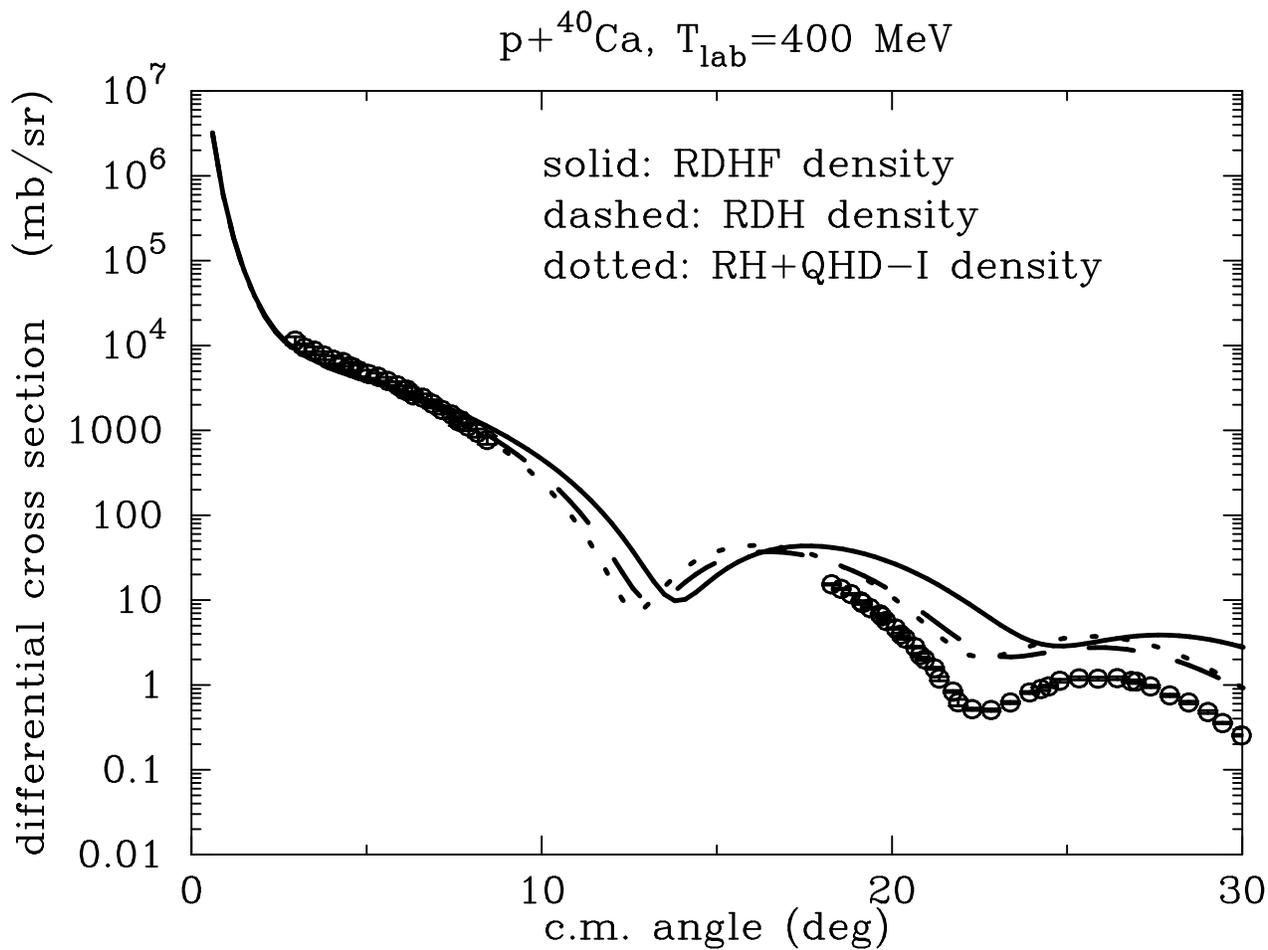
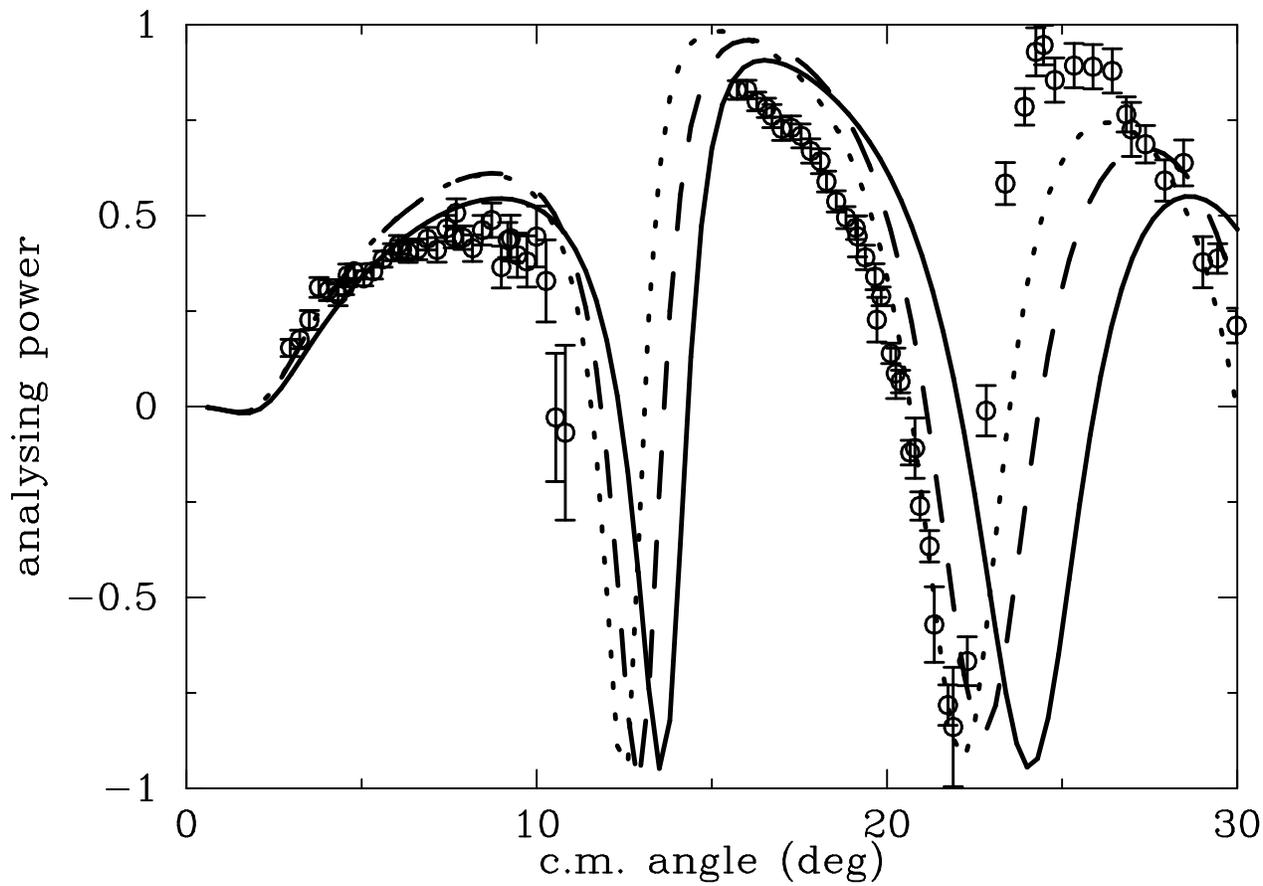
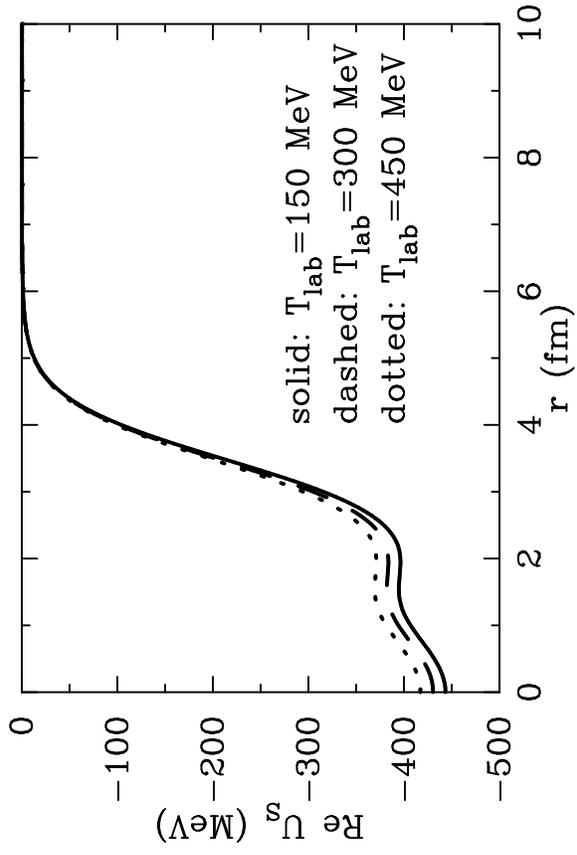
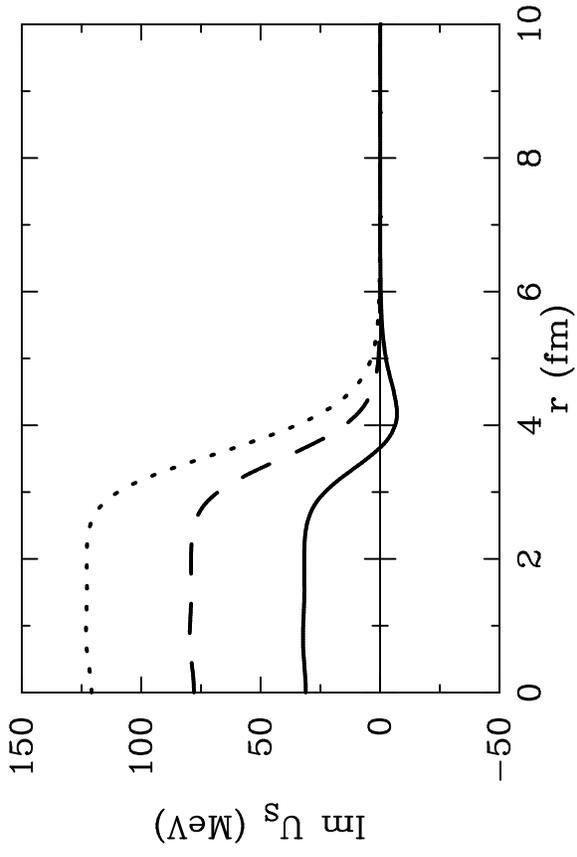
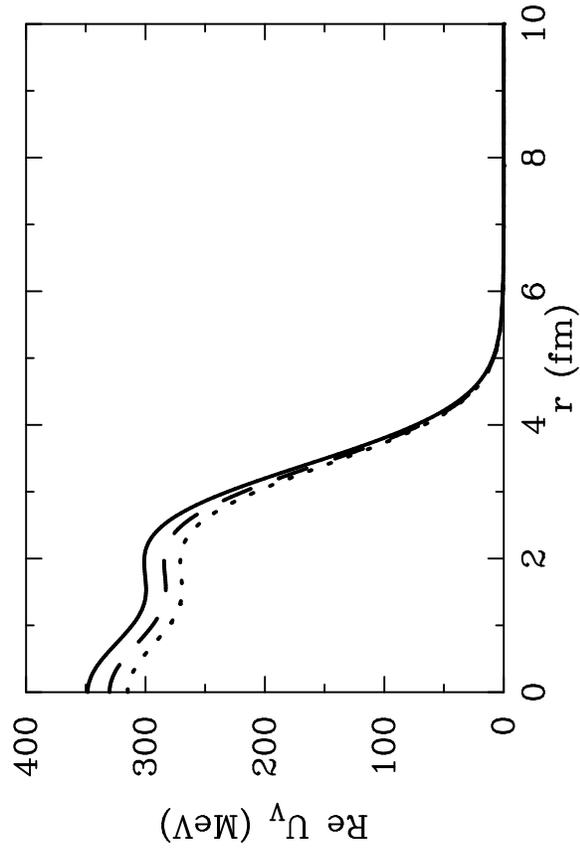
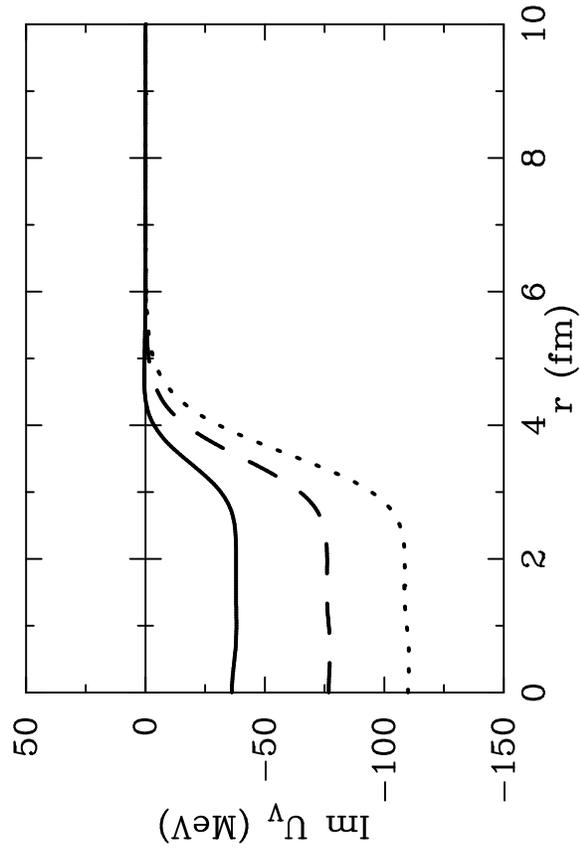
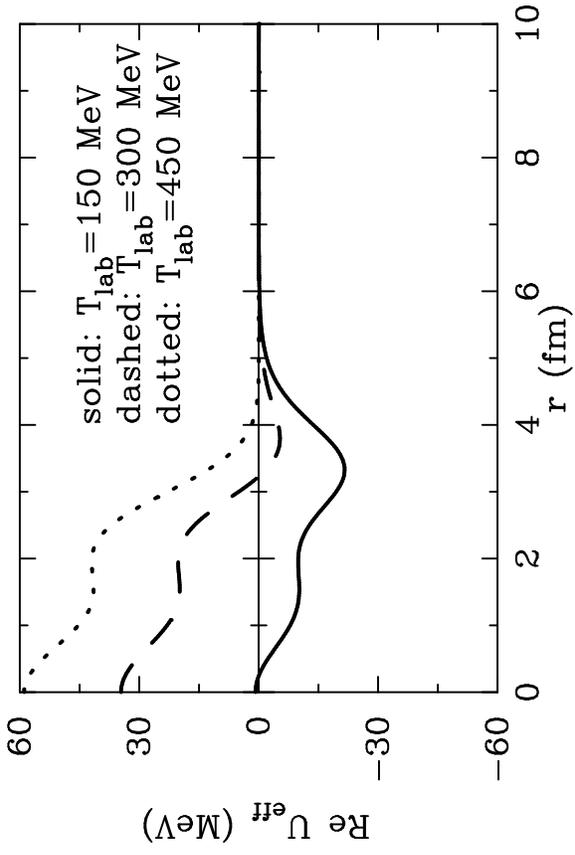
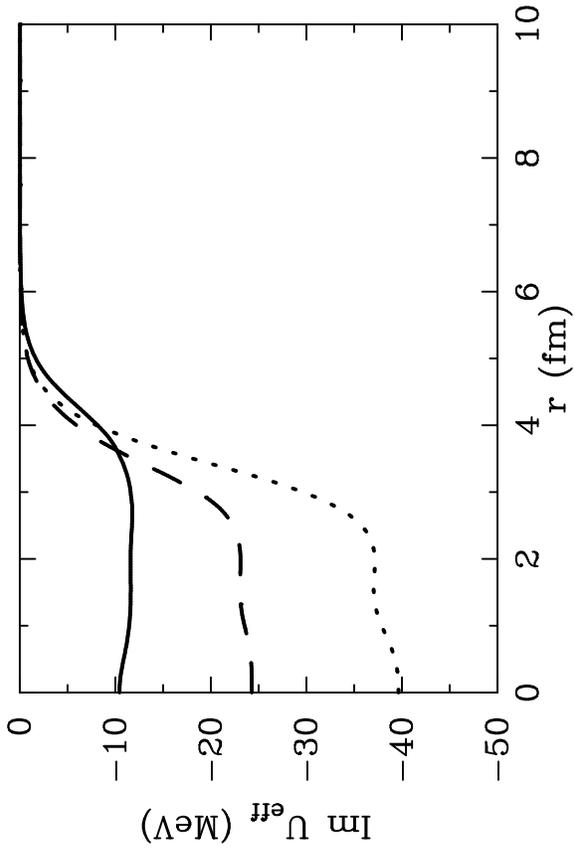
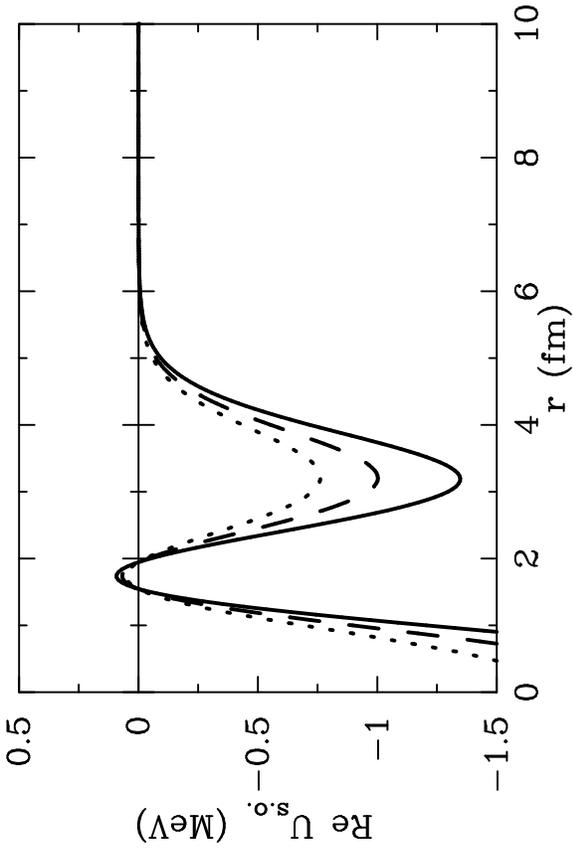
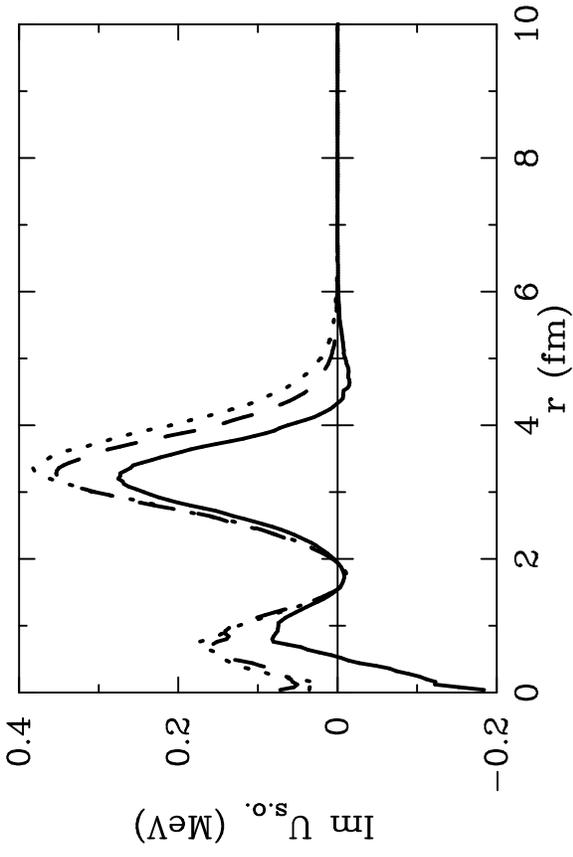
\end{document}